\documentclass[prc,twocolumn,showpacs,nofootinbib]{revtex4}
\usepackage{epsfig,amsmath,bm}
\newcommand{\mev}{\mbox{Me\hspace{-0.1em}V}}
\newcommand{\gev}{\mbox{Ge\hspace{-0.1em}V}}
\begin{document}

\title{%
$\;$\,\\[-5ex]
\hspace*{\fill}{\tt\normalsize ANL-PHY-11055-TH-2004}\\[1ex] %
Axial-vector mesons in a relativistic point-form approach}

\author{A.~Krassnigg}
\affiliation{Physics Division, Argonne National Laboratory, Argonne, IL 60439, USA}
\date{\today}

\begin{abstract}
The Poincar\'{e} invariant coupled-channel formalism for two-particle
systems interacting via one-particle exchange, which has been developed
and applied to vector mesons
in Ref.\,\cite{Kretal03a} is applied to axial vector mesons. We thereby extend
the previous study of a dynamical treatment of
the Goldstone-boson exchange by comparison with the
commonly used instantaneous approximation to the case of orbital angular
momentum $l=1$. Effects in the mass shifts show more variations
than for the vector-meson case. Results for the decay widths are sizable,
but comparison with sparse experimental data is inconclusive.
\end{abstract}


\maketitle

\section{Introduction}\label{intro}
During the past years, constituent quark models have been used successfully to
calculate spectra and properties of hadrons. In the course of the developement of these
models, relativity has emerged as a key ingredient in the light-quark sector. This
has found early consideration; e.\,g.~in the well-known works of Godfrey and Isgur \cite{GoIs85} and
Capstick and Isgur \cite{CaIs86}, where relativistic corrections were introduced in a nonrelativistic
potential model; Feynman \textit{et al.} \cite{Feetal71} based their model on a relativistic
harmonic oscillator; Carlson \textit{et al.} \cite{Caetal83} used relativistic kinetic energies
plus flux-tube motivated potential terms.
In the particular case of the Goldstone-boson-exchange (GBE) constituent-quark model
(CQM) \cite{GlRi95} the early nonrelativistic formulation \cite{Gletal96,Gletal97}
showed severe inconsistencies and was soon superseded by a ``semirelativistic''
version \cite{Gletal98}. In the semirelativistic treatment the Hamiltonian of the
model contains the relativistic kinetic energy plus potential terms, and it can in
fact be reinterpreted as the mass operator of a Poincar\'{e} invariant model as long
as the potential terms are rotationally invariant and do not depend on the total momentum
of the bound state \cite{KePo91}.
In this form the GBE CQM has been applied to spectroscopy of the light and strange
baryon sector with wide success; however, an analogous calculation for mesons
seemed to indicate failure of the model in this sector \cite{Th98}. Only
recently \cite{Kretal03a} a Poincar\'{e} invariant coupled-channel (CC) formalism for
confined two-particle systems interacting via one-particle exchange has been
shown to elucidate the relevance and applicability of the GBE CQM for vector mesons.

Aside from spectroscopy the semirelativistic GBE CQM has also been used to determine
hadronic decay widths of baryons using perturbative calculations employing elementary
emission \cite{Gletal98b,Kretal99,Pletal99} (without introducing additional parameters)
and also pair creation (one additional parameter) models \cite{Thetal00}.
At that stage a satisfactory description of experimental data was impossible.
A relativistic treatment of hadronic decays in an elementary-emission-type model
seemed natural \cite{KrKl00} and recently a Poincar\'{e} invariant decay model along the
point-form spectator approximation has been suggested \cite{Meetal02,Meetal04}. It was
observed that, in general, the theoretical results considerably underestimated the
experimental data. In the light of these results the next natural
step is to couple the decay channels explicitly to the $qqq$ (for baryons) and $\bar{q}q$ (for mesons) channels,
respectively. In \cite{Kretal03b,Kretal03c,Kretal03a} a study of vector mesons has been done to
investigate the effects of such a dynamical treatment of the exchange particle in the GBE in a Poincar\'{e}
invariant framework both in terms of mass shifts \emph{and} decay widths. The
semirelativistic form of the model produces a much too large mass splitting of $\varrho$ and $\omega$.
This flaw is removed in the CC treatment, leading to small mass shifts from the GBE
in vector mesons, which confirmes the expectation that this type of interaction should not contribute much to
the binding of such states. Regarding hadronic decays, the situation is difficult to judge with regard to comparison
with experimental data, since for only one branching ratio, which can be calculated in the model \cite{Kretal03a},
there is actually available data \cite{PDG04}.

In the present work we follow Ref.~\cite{Kretal03a}, which
contains the details of all aspects of the formalism and model used here, applying those to axial-vector
mesons with quantum numbers $J^{PC}=1^{++}$ and $J^{PC}=1^{+-}$ as well as the strange sector via
mixing of the states with the respective quantum numbers. The results identify dynamical effects for $l=1$
and enable the calculation of more hadronic decay widths within the restrictions of the states
contained in the model. We briefly review the main ingredients of the model
in Sec.~\ref{model}, the results are presented and discussed in Sec.~\ref{results},
and Sec.~\ref{conclusions} contains conclusions.

\section{Model and approximations}\label{model}
The central point of the CC treatment described in \cite{Kretal03a} is
the CC mass operator, which is defined on the little Hilbert space of the
direct sum of $\mathcal{H}_{\bar{q}q}$ (quark-antiquark) and $\mathcal{H}_{\bar{q}q\Pi}$
(quark-antiquark-pseudoscalar meson):
\begin{equation}\label{ccmassop}
M= M_c+M_I=
\begin{pmatrix}\mathcal{D}_{\bar{q}q}^c&0\\0&\mathcal{D}_{\bar{q}q\Pi}^c\end{pmatrix}
+\begin{pmatrix}0&K^\dagger\\K&0\end{pmatrix}\;.
\end{equation}
Here $M_c$ represents the diagonal part of $M$, which includes the confinement interaction
such that in both the $\bar{q}q$ and $\bar{q}q\Pi$ channels the $\bar{q}q$ pair is confined.
$M_I$ contains a vertex piece $K$ and its hermitian adjoint as defined in Eqs.~(31)
and (32) of \cite{Kretal03a}. When the eigenvalue equation
for $M$ is reduced to the $\bar{q}q$ channel, one obtains the effective interaction term
on the right-hand side of
\begin{equation}\label{theeq}
(\mathcal{D}_{\bar{q}q}^c-m)|\Psi_{\bar{q}q}\rangle=
K^\dagger(\mathcal{D}_{\bar{q}q\Pi}^c-m)^{-1}K|\Psi_{\bar{q}q}\rangle\;.
\end{equation}
In this equation, $m$ is the eigenvalue and appears also in the effective interaction term.
We note here that the interaction contains terms which correspond to the
exchange of a pseudoscalar meson $\Pi$ inside the $\bar{q}q$ pair and others, in which
the pseudoscalar meson $\Pi$ couples to the same constituent twice. We will refer to the latter
as ``loop terms''. The particular form and choices for the operators in Eq.~(\ref{theeq}) and
their matrix elements are described in detail in the appendix of \cite{Kretal03a}.
Extending the model with the parameters established in the vector-meson sector to the axial vectors,
the question for the need for readjustment of some of the parameters arises. Already for the
vector-meson calculation, the parameters used in the exchange part of the interaction were taken
without change from the semirelativistic GBE CQM in \cite{Gletal98} and we have kept them the
same here as well. This includes the definition and parameters of the quark-meson vertex form factors
which determine the range of the $\Pi$ exchange.
For the confinement piece,
a harmonic oscillator (HO) model was used in $M^2$ for two reaons: first, the HO mass operator's
eigenvalues and eigensolutions are analytically known, which facilitates the calculations
and second, it mimicks the spectrum of a linear confinement potential for $M$. The actual form
used in \cite{Kretal03a} is
\begin{equation}\label{hoev}
\mathcal{D}_{\bar{q}q}^c\rightarrow M_{nl}=\sqrt{8\;a^2\left(2n+l+3/2\right)+V_0+4\bar{m}^2}\;,
\end{equation}
where $\bar{m}^2$ is determined by the constituent quark masses, and $n$ and $l$ are the radial and
orbital angular momentum quantum numbers, respectively. $a$ is the confinement strength and $V_0$
a constant used to fix the mass of the $\varrho$ ground state to its physical value. $a$ was
adjusted such that the splitting of the $\varrho$ ground and first excited states was reproduced.
It should be noted here that this is not the best possible choice for $\mathcal{D}_{\bar{q}q}^c$ in terms of an
accurate fit to experimental data for higher excited states. However, the main emphasis of our
studies still lies on the effects of a dynamical treatment of the one-particle exchange as compared
to an instantaneous approximation (IA). Therefore $\mathcal{D}_{\bar{q}q}^c$ is sufficient for the
present purpose. In principle one could choose any confinement
operator, which is diagonal in the basis of Eq.~(\ref{theeq}), satisfies the (in our case point-form)
Bakamjian-Thomas requirements \cite{KePo91},
and whose solutions are known and can therefore be used to discretize the problem.

In an attempt to extend the basic HO piece of the model beyond the vector-meson sector one
can use the concept of Regge trajectories \cite{TaNo00}. This approach has already been used
in \cite{Feetal71} and recently in the context
of relativistic Hamiltonian dynamics \cite{AnSe99}. If one uses such a trajectory, which
contains the $\varrho$ ground state, to determine the parameters $a$ and $V_0$, one finds that the
parameter set established for the vector mesons does not need to be changed. The parameters used
in all calculations presented here are thus $a=312$ \mev\  and $V_0=-1.04115$ \gev. This completes
the summary of the model definitions.

In the calculations we make two approximations. First, we do not treat the loop contributions
in the effective interaction explicitly. This is motivated by the assumption that their effects can
be accounted for via a change in the constituent-quark mass. An explicit treatment of these terms
is an extension of the model which will be incorporated in future studies. Second, some of
the matrix elements occurring in Eq.~(\ref{theeq}) contain Wigner rotations, which come from the
overlap of the various sets of basis states used in the computation of the effective interaction
(see Eq.~(A16) in \cite{Kretal03a}).
We neglect these rotations, because, while the numerical effort to include them is considerable,
their effects have been found to be small compared to boost effects in calculations of
electromagnetic form factors of the nucleon \cite{Waetal00}.

\section{Results and discussion}\label{results}
\begin{figure*}
\epsfig{file=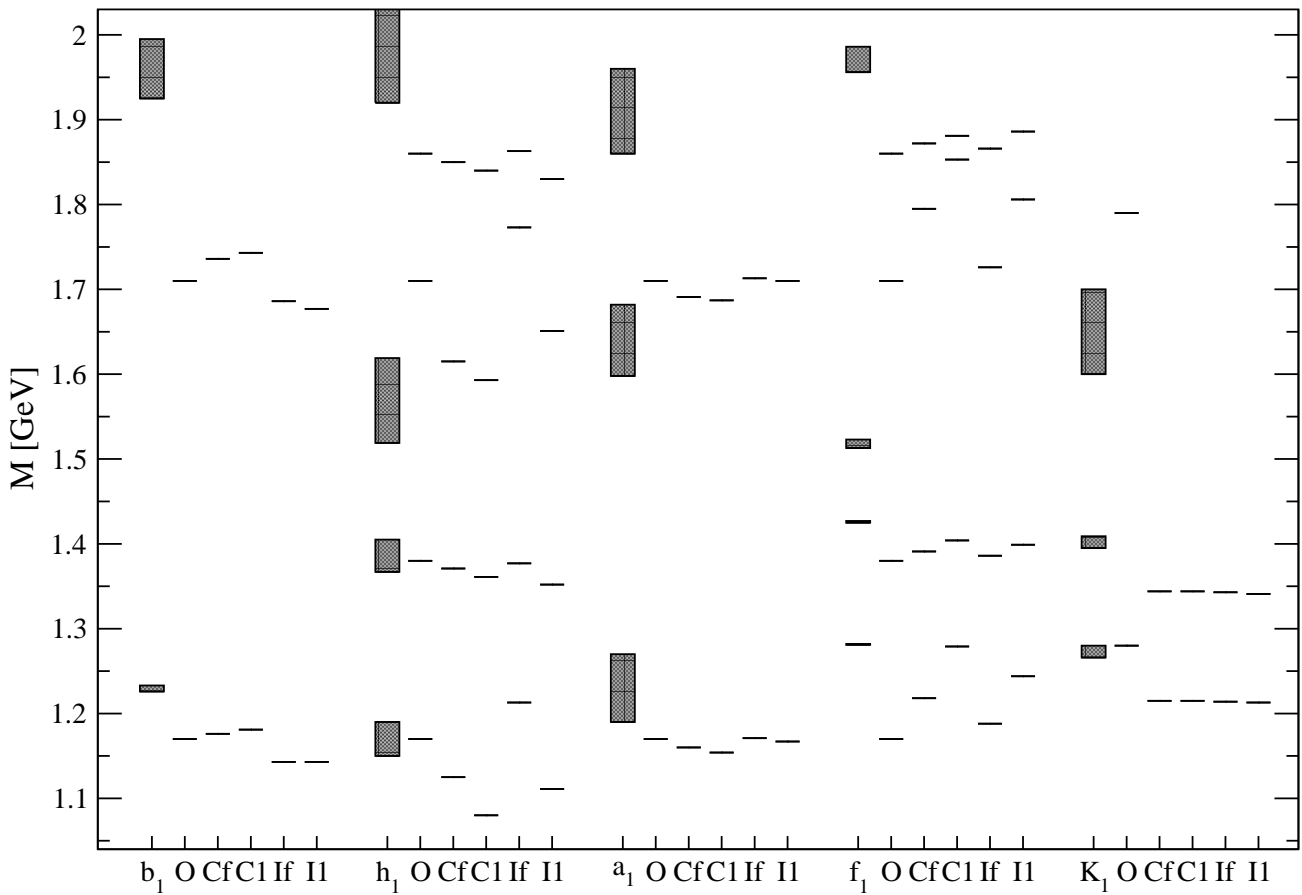,width=12cm,clip=true,angle=270}
\caption{The results for axial-vector meson spectra: Experimental data with uncertainties (particle
name), oscillator (O), CC with vertex form factor (Cf), CC with FF=1 (C1), instantaneous approximation
with vertex FF (If), and IA with FF=1 (I1)\label{spectra}}
\end{figure*}
We have obtained results for the axial-vector states with quantum numbers $J^{PC}=1^{++}$ and
$J^{PC}=1^{+-}$ by solving the eigenvalue equation for $M$ numerically.
For quark-model mesons one has the relations $P=(-1)^{l+1}$ and $C=(-1)^{l+s}$, where $l$ is
the orbital angular momentum and $s$ the total spin of the constituents, respectively; for
$J^{PC}=1^{++}$ this entails $s=1$, $l=1$; the physical states corresponding to
this set of quantum numbers are the $f_1$ (with isospin $I=0$), $a_1$ ($I=1$), and $K_{1A}$ ($I=1/2$).
for $J^{PC}=1^{+-}$ one gets $s=0$, $l=1$ with the associated particles  $h_1$ ($I=0$),
$b_1$ ($I=1$), and $K_{1B}$ ($I=1/2$). Within the isospin 0 channels we assume ideal mixing between
the octet and singlet $SU(3)$-flavor configurations, meaning that the $h_1$ as well as the $f_1$
spectra each contain both pure $\bar{n}n$ and $\bar{s}s$ states (in the usual notation,
$n$ here denotes light quarks).
In the strange sector, the physical states of the $K_1$ spectrum are mixtures of the $K_{1A}$
and $K_{1B}$, since they are not charge-parity eigenstates. In our treatment of this mixing we
follow Blundell \textit{et al.} \cite{Bletal95}.
The results are presented in Fig.~\ref{spectra} in six ``columns'' for each set of quantum numbers;
the experimental values \cite{PDG04} are depicted in the first column (denoted by the particle
names) by boxes indicating the experimental
uncertainties; the second to sixth columns contain results for pure oscillator (O), coupled channel
calculation with vertex form factor (Cf), CC calculation with the form factor set =1 (C1),
an instantaneous approximation with vertex form factor (If), and the IA calculation with the form
factor set =1 (I1). These are the same categories as presented in Ref.~\cite{Kretal03a}; also see this
reference for details.

In the vector-meson sector \cite{Kretal03a} the $\bar{q}q$ states have mainly orbital angular momentum
$l=0$ with small admixtures of $l=2$. There the main observation was that generally the CC treatment
produces smaller mass shifts than the IA, including the prominent case
of the $\omega$ ground state. For axial-vector mesons one always has $l=1$ and the observations about
mass shifts are different, except that the main differences between CC treatment and IA can be
found in the isoscalar channels; this is not surprising, because one-pion exchange is strongest
in these channels and the light mass of the exchange particle plays a central role in the dynamical
setup.

In general the dependence of the mass shifts on the use of a form factor at the quark-meson vertex
is smaller in the CC treament than in the IA. This is true in particular for
the case of the $h_1$ meson, where for the IA the shifts with and without the form factor have
opposite sign. For the $b_1$, a different sign change appears: while the CC shifts are positive,
the IA ones are negative (although in both cases the shifts are small). In the $f_1$ spectrum,
one observes that mass shifts using the form factor are larger for the CC than the IA results and
the same seems to apply to the results without the form factor except for the first $\bar{s}s$
state. This is opposite to the general observation for vector mesons. The reason for these
variety of effects lies in the complexity of the dynamical setup for the CC formalism in
connection with the wave functions for $l=1$ states. On different ranges of the relative
momentum between the $q$ and the $\bar{q}$, these can have support of different sign, which
gets modified in addition via the kinematical relations used in the calculation as apparent
from Eq.~(A16) in \cite{Kretal03a}.

The results for the decay widths are larger than
in the vector-meson case. However, similarly to the latter there is only one branching ratio
in \cite{PDG04} which we can compare our results to, although most of the decays are regarded as ``seen''.
Our results are generally of the order of $\approx 20$ to $90$ \mev. The known branching ratio is that of the
$b_1(1235)$ with $\Gamma=142\pm9$ \mev, which decays dominantly into channels contained in our model;
our result is 33 \mev\ with and 42 \mev\ without the form factor, underestimating experiment by about
a factor of 4.

\section{Conclusions}\label{conclusions}
We have applied the Poincar\'{e} invariant coupled-channel (CC) formalism of Ref.\,\cite{Kretal03a}
to the sector of axial-vector mesons. The axial-vector quantum numbers $J^{PC}=1^{++}/1^{+-}$ both
imply orbital angular momentum $l=1$ for the $\bar{q}q$ pair. The effects
of the dynamical treatment of the one-boson exchange (OBE) as compared to the instantaneous approximation (IA)
reveals different characteristics as compared to the case of vector mesons (mainly $l=0$). There
are four main observations: i) as for the vector mesons, the effects are strongest in the isoscalar
channels, since there one-pion exchange dominates, which is very sensitive to the dynamical setup
due to the light pion mass. ii) in the $h_1$ spectrum the use of a vertex form factor (as compared
to a form factor =1) changes the sign of the mass shift from the meson exchange in the IA,
while the CC results have the same sign and magnitude regardless of
the details of the form factor. iii) in the $f_1$ spectrum the CC mass shifts are larger than the
IA ones - opposite to the general trend (including the vector mesons). iv) in the $a_1$ spectrum
the CC shifts have the opposite sign as compared to the IA ones. The main conclusion from this
collection of observations must be that results from a dynamical CC treatment of one-particle exchange
can differ significantly, both in magnitude and sign, in channels where this exchange is
important. The results for the decay widths are sizable, but comparison with experimental data is
inconclusive: the only data point with definite value (for the $b_1(1235)$) is underestimated by
a factor of 4. This supports the conclusion drawn in \cite{Kretal03a} from an
analogous situation in the $\omega(1420)$ case: the calculations could be improved by explicitly
including the loop contributions from the interaction in Eq.~(\ref{theeq}) and/or taking into
account final-state interactions.

These conclusions strongly suggest analogous investigations of $qqq$ systems in this context, since
such a dynamical, Poincar\'{e} invariant treatment of OBE in the baryon sector is yet
missing. In \cite{Kretal03a} and the present work the path is laid out and also a possible
intermediate step has been identified \cite{Kletal01}.
We note here that a treatment along the lines of the stochastic variational method \cite{VaSu95}
used up to date in the GBE CQM \cite{Gletal98} seems impossible due to the high
dimensions and numerical nature of the integrations involved in the solution of the CC problem.
A more promising approach is of Faddeev type along the lines of Ref.~\cite{Paetal00a}.
It is important to proceed in this direction, because:
a dynamical treatment of the OBE in baryons will yield hadronic baryon decay widths
in a nonperturbative way, which could remedy their unsatisfactory description at the
present stage of the model. Furthermore, given the limited comparison to experimental
data of hadronic meson decays predicted by the present work and in Ref.\,\cite{Kretal03a} as well as the
importance of GBE in the baryon sector,
an analogous investigation of baryons will make the full impact of a dynamical treatment of
OBE-type interactions clear.

\begin{acknowledgments}
We acknowledge valuable discussions with F.\,Coester, W.\,H.\,Klink, T.\,Melde, and W.\,Schweiger.
This work was supported by: \textit{FWF} via
\textit{Erwin-Schr\"odinger-Stipendium} no.\ J2233-N08 and the Department of Energy,
Office of Nuclear Physics, contract no.\ W-31-109-ENG-38.
\end{acknowledgments}

\end{document}